# A Quantum Cognition Analysis of Human Behaviour by Hardy's Non-locality Argument

Pegah Imannezhad[1] and Ali Ahanj[*,1]

[1] *Department of Physics, Khayyam University, Mashhad, Iran.*

**Abstract**

Quantum cognition is an emerging field making uses of quantum theory to model cognitive phenomena which cannot be explained by classical theories. Usually, in cognitive tests, subjects are asked to give a response to a question, but, in this paper, we just observed the subjects' behaviour and the question and answer method was not applied in order to prevent any mental background on participants' minds. Finally, we examined the experimental data on Hardy's non-locality argument (HNA), and we noticed the violation of HNA in human behaviour.

## 1 Introduction

Quantum mechanics was created to explain the puzzling findings that were impossible to understand by using classical theories [1, 2, 3]. Previously, all cognitive researches relied on classical probability theory together with principles of classical mechanics. However, it has recently been found that some experimental data on human cognition such as the violation of sure-thing principle [4], conjunction fallacies [5], disjunction fallacies [6] and order effects [7, 8] cannot be studied via classical theory.

In recent years, many researches have been done on using quantum theory in cognitive science [9, 10]. Research in the application of quantum theory in cognition science has created a new cognitive field called quantum cognition. According to some aspects of quantum theory such as complementarity, superposition, contextuality and entanglement, quantum probability seems to be a useful framework to describe a wide variety of subjects' behaviour.

We know the brain is a complex state of biological material; how it works and interacts with the external environment still is not well and fully understood [11, 12]. So subjects' behaviour is often claimed to be irrational [13]. There have been a number of theories about the brain function throughout the years. For some, the brain is a complex neural network obeying classical theories [14, 15, 16]; on the other hand, some researchers believe non-classical effects are responsible for brain function [17, 18, 19, 20]. It is obvious that the interaction between brain and environment needs to be examined thoroughly.

---

[*] a.ahanj@khayyam.ac.ir; ahanj@ipm.ir

The conceptual combination is an important concept in the field of cognitive science for understanding how concepts are combined among human brains. Aerts and Sozzo studied the combination of two concepts, the concept of Animal, and the concept of Acts, in the sentence, "The Animal Acts" [21]. The collected data showed the violation of Bell's inequalities [22], and it revealed the identification of entanglement in concept combination.

In some papers, it is asserted that cognitive experiments are not as direct as experiments in quantum mechanics [23, 24]. In fact, in almost all cognitive experiments, subjects are required to give an answer to a question, and they usually provide a response that they think is correct, and not the first answer that crosses their minds; this means that the subjects' answers are stochastic, but they do not give a random answer like quantum particles. Therefore, in this research, we have studied the behaviour of sales operators and their customers of a well-equipped call centre[2], solely by observing the subjects' behaviour without the question and answer method. In fact, we investigated the subjects' decisions so as to apply the maximum possible accuracy to the data collection.

As we know Bell's inequality is not the only way to express non-locality in quantum mechanics. Lucien Hardy describes the equations which show the quantum contradicts directly with local realism (not with inequalities such as Bell's inequality) [25, 26]. Thus, in the present study, we examined the experimentally collected data on Hardy's non-locality argument (HNA) to check whether the collected data follows the classical theory or not. In the following lines, first, we briefly discuss the theoretical basis of HNA in Sec.2. Then in Sec.3, we explicitly describe the experiment that we have already done, and, in Sec.4, we examine the obtained results on HNA (shown in Appendix); moreover, in Sec.5, we arrive at the conclusions of this research.

## 2 Hardy's non-locality argument (HNA)

Lucien Hardy provided an argument which revealed non-locality within quantum mechanics. Hardy's non-locality was later expanded by Cabello[27].

The Hardy's logical structure is as follows: consider four events $a_1, a_2, b_1$ and $b_2 \in \{+1, -1\}$, where the positive sign means that an event occurs and the negative sign means that the event does not take place. Also, we consider two observers, Alice and Bob, where $a_1$ and $a_2$ may happen on Alice's side and $b_1$ and $b_2$ may happen on Bob's side which is far apart from Alice. We can represent Hardy's non-locality argument (HNA) as:

$$Pr(a_1 = +1, b_1 = +1) = 0 \quad (1)$$
$$Pr(a_1 = -1, b_2 = +1) = 0 \quad (2)$$
$$Pr(a_2 = +1, b_1 = -1) = 0 \quad (3)$$
$$Pr(a_2 = +1, b_2 = +1) = q \quad (4)$$

---

[2] Yas Sefid Pars's call center

The probability $Pr(a_1 = +1, b_1 = +1) = 0$ means that, if $a_1$ is measured on Alice's side and $b_1$ is measured on Bob's side, then the probability that both will get +1 is zero. Other probabilities can be analysed in a similar way. The equations (1)-(4) form the basis of HNA.

It can be easily seen that these equations contract local-realism if $q \neq 0$. In the local-realism, the value of $q$ is zero, but in the quantum mechanics, for the non-maximum entangled state, the value of $q$ is more than zero, and the upper bound of $q$ in quantum mechanics is 0.09[28].

## 3 Description of the Experiment

In this study, we just observed the behaviour of the sales operators and their customers through the incoming calls, so-called inbound calls. In the call centre, where we did our experiment, the number of inbound calls was more than 2000 per working day and more than 1200 calls per holiday. The average number of inbound operators was 19 people per working day and 13 people per holiday.

As for surveying the research results by HNA, we assigned four events $a_1, a_2, b_1$ and $b_2$ as follows. We considered the parameter $a_1$ as an abandoned call (An abandoned call is a call that is ended before any conversation occurs). It usually happens when there is a waiting queue and the caller is frustrated with the time on hold; therefore, once a customer ended her/his call before an operator's response, $a_1 = +1$, and when her/his call was being answered by an operator, $a_1 = -1$. Also, we considered $a_2$ as a purchase, thus $a_2 = +1$ the moment a customer purchased a service, and $a_2 = -1$ when she/he did not purchase any services (In the call centre, operators were selling Internet services). We considered $b_1$ when an operator was responding to the incoming calls, thus $b_1 = +1$ as long as the operator was responding, and $b_1 = -1$ when she/he was not responding. It should be noted that the operators were allowed to change their panels' status to DND[3] mode for the purpose of resting. As a result, during the time, when their panels were in DND mode or there were not any incoming calls, the operators were at work, but they were not responding. Finally, we took $b_2$ when it was not an operator's working time, so as long as the operator was not at work, $b_2 = +1$ and when she/he was present at work, $b_2 = -1$. As we notice, events $a_1$ and $a_2$ are related to the customers' behaviour and events $b_1$ and $b_2$ are related to the operators' behaviour.

According to the above-mentioned data, the Eq. (1) is the probability that an operator was responding to a disconnected call, and the Eq. (2) is the probability that a customer's call was being responded by an operator who was not at work. The Eq. (3) is the probability that a customer purchased a service when none of the operators was responding to her/his call. Finally, the Eq. (4) is the probability that a customer purchased a service, and this sale was registered for an operator who was not at work at the time of purchase.

---

[3] Do Not Disturb

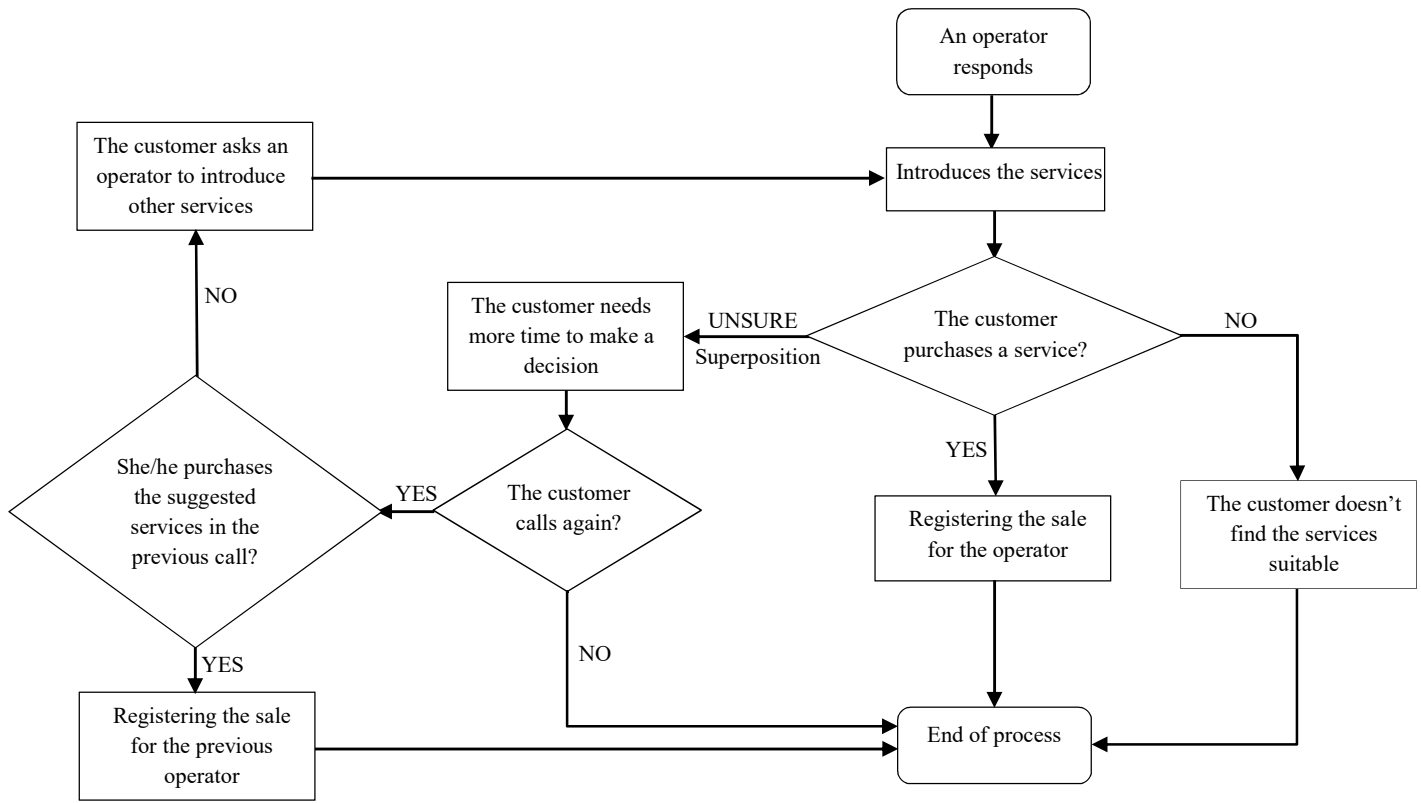

Fig. 1: The sales process flow diagram

The sales process in the call centre was in this fashion. After an operator responded to a customer's call, the operator would suggest a service to the customer according to the customer's needs. If the customer purchased the service, the sale would be registered for the operator. But sometimes, the customer remained confused about purchasing the service or not, and she/he needed more time to make a decision. In this situation, whenever the customer made a decision and called again, the sale would be registered for the first operator (the operator who answered the customer's call at the first time) and the second operator (the operator who answered the customer's call at the second time) just registered the sale, if the client purchased the service that was suggested in the previous call. On the other hand, if the customer asked the second operator to introduce other services and the customer purchased the service that was suggested in her/his second call, the sale would be registered for the second operator. In fact, each sale would be registered for the operator who convinced the client to purchase the sold service.

To collect the customers' information and avoiding disruption in allocating commission to the operators, all the responding and sales processes were registered in the call centre database. Also, all the conversations between the operators and their customers were recorded too. The sales process flow diagram is shown in Fig. 1.

## 4 Results and Discussion

The results that are shown in Appendix include the number of calls responded by the operators, the number of abandoned calls, the amount of sales made by operators who were not at work at the time of purchase (absent operators) and the amount of sales made by operators who were at work at the time of purchase (present operators) concerning all calls from August 22, 2016 to October 21, 2016. Totally, over 115000 calls have been reviewed in this study.

Returning to HNA equations (Eq.(1)-Eq.(4)), we know, once a customer ended her/his call before an operator responded, there was no chance for the operator to answer the abandoned call, so the Eq.(1) is zero, and when it was not an operator's working time, she/he could not respond to the incoming calls, so again, the Eq.(2) is zero. Moreover, while no-one was responding to a customer's call, she/he could not purchase any services, thus the Eq. (3) is zero as well.

Finally, it is expected that when it was not during an operator's working time, she/he was not able to sell because she/he was not present at the call centre to respond to the incoming calls and encourage the customers to purchase, but the collected data (column 4 of the table in Appendix) shows the registered sales for the absent operators.

We know that, after hearing an operator's presentation, the customer must decide whether to purchase the service or not, but she/he may be indecisive which allows both of these definite states (purchasing or not purchasing) to have the potential for being expressed at any given moment. In fact, as long as the customer is indecisive, she/he is in a superposition state that makes her/him feel confused, or uncertain. Sometimes the customer makes a decision about purchasing the suggested service after working time of the operator who presented the service to the client. That is why we observed the sales records made in absence of operators.

Consequently, the Eq. (4) is equal to the total sales amounts made in absence of operators (the total sum of column 4 of the table) divided by the total sales amounts made by all the operators (the total sum of columns 4 and 5 of the table).

$$Pr(a_2 = +1, b_2 = +1) = \frac{50373989}{1273102156 + 50373989} = 0.038062$$

The amount obtained shows that in this experiment the human behaviour is not in conformity with local realism.

## 5 Conclusions

In most of the cognitive experiments, the question and answer method is used. In this case, there is the likelihood that subjects' mental background, environment and other external factors might affect the subjects' response. In other words, subjects initially evaluate questions in different sample spaces in their minds and then express their answers, while researchers are not aware of those sample spaces. Therefore, we solely observed the subjects' behaviour (the sales operators' and their customers') for collecting data with the highest possible accuracy.

According to the collected data in this experiment, subjects did not behave classically. Therefore, it is recommended that in the research studies related to the analysis of individuals' behaviour, not only the results should be checked in classical mechanics, but also they are expected to be studied in quantum mechanics.

## 6 Acknowledgments

This research was partially supported by Yas Sefid Pars Co. We thank Dr K. Javidan and Dr V. Salary for providing insight and expertise that greatly assisted the research. Also, we thank Dr O. Ghahraman for his comments that improved the manuscript.

## 7 Appendix

The results of this review are presented in the following table which includes the date, the number of calls that were responded by the operators, the number of abandoned calls, the sales amount of the absent operators and the sales amount of the present operators, regarding all calls from August 22, 2016 to October 21, 2016.

| Date | The number of responded calls | The number of abandoned calls | The sales amount of the absent operators (Toman) | The sales amount of the present operators (Toman) |
|---|---|---|---|---|
| 22-Aug-16 | 2473 | 164 | 874,180 | 18,654,118 |
| 23-Aug-16 | 2288 | 151 | 886,860 | 25,890,741 |
| 24-Aug-16 | 2112 | 54 | 751,010 | 23,879,117 |
| 25-Aug-16 | 1919 | 67 | 717,220 | 23,697,522 |
| 26-Aug-16 | 1300 | 23 | 476,330 | 14,345,285 |
| 27-Aug-16* | 1146 | 44 | 725,286 | 26,122,732 |
| 28-Aug-16 | 2072 | 63 | 478,728 | 23,064,912 |
| 29-Aug-16 | 2056 | 51 | 1,054,030 | 22,061,473 |
| 30-Aug-16 | 1984 | 41 | 725,286 | 24,230,032 |
| 31-Aug-16 | 1929 | 48 | 998,974 | 20,948,317 |
| 01-Sep-16 | 1815 | 90 | 825,784 | 24,046,560 |
| 02-Sep-16* | 581 | 38 | 961,380 | 14,216,797 |
| 03-Sep-16 | 2282 | 77 | 634,660 | 26,390,214 |
| 04-Sep-16 | 2022 | 104 | 542,820 | 24,551,341 |
| 05-Sep-16 | 2030 | 53 | 877,050 | 23,821,728 |
| 06-Sep-16 | 1888 | 70 | 574,430 | 22,233,120 |
| 07-Sep-16 | 1964 | 66 | 518,460 | 21,551,614 |
| 08-Sep-16 | 1723 | 122 | 913,585 | 21,202,381 |
| 09-Sep-16 | 1046 | 17 | 609,310 | 12,161,949 |

| Date | | | | |
|---|---|---|---|---|
| 10-Sep-16 | 2121 | 109 | 471,970 | 24,632,378 |
| 11-Sep-16 | 1885 | 68 | 554,810 | 20,242,018 |
| 12-Sep-16 | 1230 | 38 | 852,380 | 12,293,565 |
| 13-Sep-16 | 1983 | 64 | 909,060 | 24,122,251 |
| 14-Sep-16 | 1969 | 78 | 994,080 | 22,803,851 |
| 15-Sep-16 | 1743 | 45 | 373,870 | 18,951,874 |
| 16-Sep-16 | 1208 | 76 | 1,171,750 | 12,741,206 |
| 17-Sep-16 | 2206 | 168 | 540,095 | 26,098,073 |
| 18-Sep-16 | 2087 | 53 | 192,930 | 25,767,299 |
| 19-Sep-16 | 1931 | 62 | 733,570 | 22,546,980 |
| 20-Sep-16 | 1327 | 79 | 1,025,690 | 14,204,147 |
| 21-Sep-16 | 2201 | 183 | 734,660 | 27,394,389 |
| 22-Sep-16 | 1947 | 58 | 744,470 | 24,257,760 |
| 23-Sep-16 | 1187 | 32 | 1,020,240 | 12,815,533 |
| 24-Sep-16 | 2040 | 124 | 883,990 | 25,229,192 |
| 25-Sep-16 | 1993 | 80 | 141,700 | 25,633,633 |
| 26-Sep-16 | 1136 | 62 | 813,850 | 24,830,220 |
| 27-Sep-16 | 2103 | 88 | 960,290 | 23,767,149 |
| 28-Sep-16 | 1963 | 88 | 954,840 | 22,621,980 |
| 29-Sep-16 | 1838 | 24 | 1,230,610 | 22,002,223 |
| 30-Sep-16 | 1211 | 50 | 995,170 | 11,701,793 |
| 01-Oct-16 | 2212 | 122 | 832,760 | 22,243,989 |
| 02-Oct-16 | 1997 | 60 | 779,350 | 22,683,902 |
| 03-Oct-16 | 1992 | 70 | 1,111,000 | 23,176,102 |
| 04-Oct-16 | 1969 | 179 | 927,045 | 22,478,785 |
| 05-Oct-16 | 1996 | 75 | 522,873 | 20,203,281 |
| 06-Oct-16 | 1644 | 126 | 922,503 | 19,289,878 |
| 07-Oct-16 | 1319 | 43 | 1,021,330 | 12,365,875 |
| 08-Oct-16 | 2055 | 117 | 1,216,440 | 22,497,115 |
| 09-Oct-16 | 1892 | 115 | 679,070 | 20,037,633 |
| 10-Oct-16 | 1978 | 145 | 1,158,670 | 19,257,188 |
| 11-Oct-16 | 977 | 30 | 246,340 | 9,233,706 |
| 12-Oct-16* | 758 | 9 | 360,790 | 8,030,647 |
| 13-Oct-16 | 2132 | 83 | 1,334,705 | 23,287,878 |
| 14-Oct-16 | 1497 | 57 | 991,900 | 15,651,357 |
| 15-Oct-16 | 2668 | 172 | 1,016,570 | 27,203,122 |
| 16-Oct-16 | 2276 | 127 | 832,595 | 23,660,962 |
| 17-Oct-16 | 1219 | 104 | 1,273,120 | 24,260,629 |
| 18-Oct-16 | 2341 | 276 | 1,115,070 | 23,857,171 |
| 19-Oct-16 | 2186 | 118 | 790,250 | 20,509,051 |
| 20-Oct-16 | 2092 | 125 | 1,593,930 | 20,167,315 |
| 21-Oct-16 | 1453 | 61 | 1,202,270 | 15,279,103 |
| **SUM** | **110592** | **5186** | **50,373,989** | **1,273,102,156** |

Table 1: The collected data from August 22 to October 21. *It should be noted that on August 27, September 2 and October 12, there were interruptions in communications. Also notice that Toman is Iranian currency.